\documentclass[pdftex,aps,prl,twocolumn,showpacs,amsmath,floatfix]{revtex4}
\pdfoutput=1
\usepackage[latin1]{inputenc}
\usepackage{latexsym,amstext,amssymb,amsmath}
\usepackage[dvips]{graphicx}
\usepackage{color}
\usepackage{verbatim}

\newcommand{\cm}{cm$^{-1}$}

\begin{document}

\title{
Chemical Raman Enhancement of Organic Adsorbates on Metal Surfaces
}

\author{A. T. Zayak$^{1,2}$}

\author{Y. S.  Hu$^{3}$}

\author{H. Choo$^{1,2}$}

\author{J. Bokor$^{1,2}$}

\author{S. Cabrini$^{1}$}

\author{P. J. Schuck$^{1}$}

\author{J. B. Neaton$^{1}$}
\email[Electronic address:]{jbneaton@lbl.gov}

\affiliation{$^{1}$ Molecular Foundry,
  Lawrence Berkeley National Laboratory, Berkeley  CA 94720, USA}

\affiliation{$^{2}$ Department of Electrical Engineering and Computer
  Sciences, UC Berkeley, Berkeley CA  94720-1770, USA}

\affiliation{$^{3}$ Bioengineering Department, Rice University, Houston, TX
  77005 }

\date{\today}

\begin{abstract}

Using a combination of first-principles theory and experiments, we
provide a quantitative explanation for chemical contributions to
surface-enhanced Raman spectroscopy for a well-studied organic
molecule, benzene thiol, chemisorbed on planar Au(111) surfaces. With
density functional theory calculations of the static Raman tensor, we
demonstrate and quantify a strong mode-dependent modification of
benzene thiol Raman spectra by Au substrates. Raman active modes with the largest
enhancements result from stronger contributions from Au to their
electron-vibron coupling, as quantified through a deformation
potential, a well-defined property of each vibrational mode. A
straightforward and general analysis is introduced that allows
extraction of chemical enhancement from experiments for specific
vibrational modes; measured values are in excellent agreement with our
calculations.

\end{abstract}


\maketitle

\marginparwidth 2.7in

\marginparsep 0.5in


The ability to detect and characterize chemical species at the
single-molecule level requires probes at the limits of present
experimental resolution, and is a fundamental challenge to
nanoscience. Since its discovery over three decades ago, surface
enhanced Raman spectroscopy (SERS) has shown significant promise for
sensing individual molecules adsorbed near metal nanostructures or
substrates with nanoscale roughness\cite{Fleischman,Jeanmaire,MGAlbrecht}. In SERS, the conversion of
incident light into surface plasmons near asperities on metal
surfaces, combined with chemical and resonant effects, has been
reported to yield Raman cross sections increased by factors of up to
10$^{14}$, enabling single-molecule detection \cite{Moskovits,Kneipp,Brus2000,Nie,Haran,Natelson}. While the enhancement
associated with surface plasmons can reach 10$^8$ \cite{Willets}, 
remaining increases and associated changes in mode frequencies
have been reported to arise from chemical adsorption
\cite{MoskovitsRev,Campion1998,Moskovits,Jensen2008,Morton,Heller,Persson},
as well 
as resonant intra-molecular and metal-molecule charge transfer \cite{Adrian,Arenas,Lombardi2008}.
Despite a wealth of prior theoretical studies, these chemical
enhancement (CE) mechanisms have remained poorly understood and
difficult to quantify \cite{Campion1998,Moskovits,Jensen2008,Morton}.  
While for some vibrational modes CE is estimated \cite{MoskovitsRev,Campion1998,Moskovits,Maitani2009}
to be about 10-100, there is currently no clear picture for why 
certain modes are enhanced more than others. Previous theoretical
studies have proposed mechanisms for this mode dependence \cite{Heller,Persson,Morton,Adrian,Arenas,Lombardi2008}, but
none of these models have been validated by experiments or more
rigorous first-principles calculations, leaving the origin of dominant
chemical contributions to SERS an open question \cite{Campion1998,Moskovits,Jensen2008,Morton}. 

In this Letter, we use a combination of first-principles calculations
and experiment to demonstrate the origin of chemical contributions to
SERS for benzene thiol (BT) molecules chemisorbed to Au surfaces. From
our density functional theory (DFT) calculations of static
contributions to the Raman tensor, we elucidate the vibrational mode
dependence \cite{Campion1998,Moskovits,Jensen2008,Morton} of the chemical enhancement, explicitly relating modes
with the largest CE to those with the greatest mode-induced shift of
the molecular frontier orbital energy, as quantified through a
deformation potential. Relative CE of BT vibrational modes measured at
different probe frequencies agree quantitatively with our static
calculations for all binding sites considered. While the magnitude of
CE for BT on Au is sensitive to binding motif, its relative value is
not. 
%
\begin{figure}[h]
\includegraphics*[width=7.0cm]{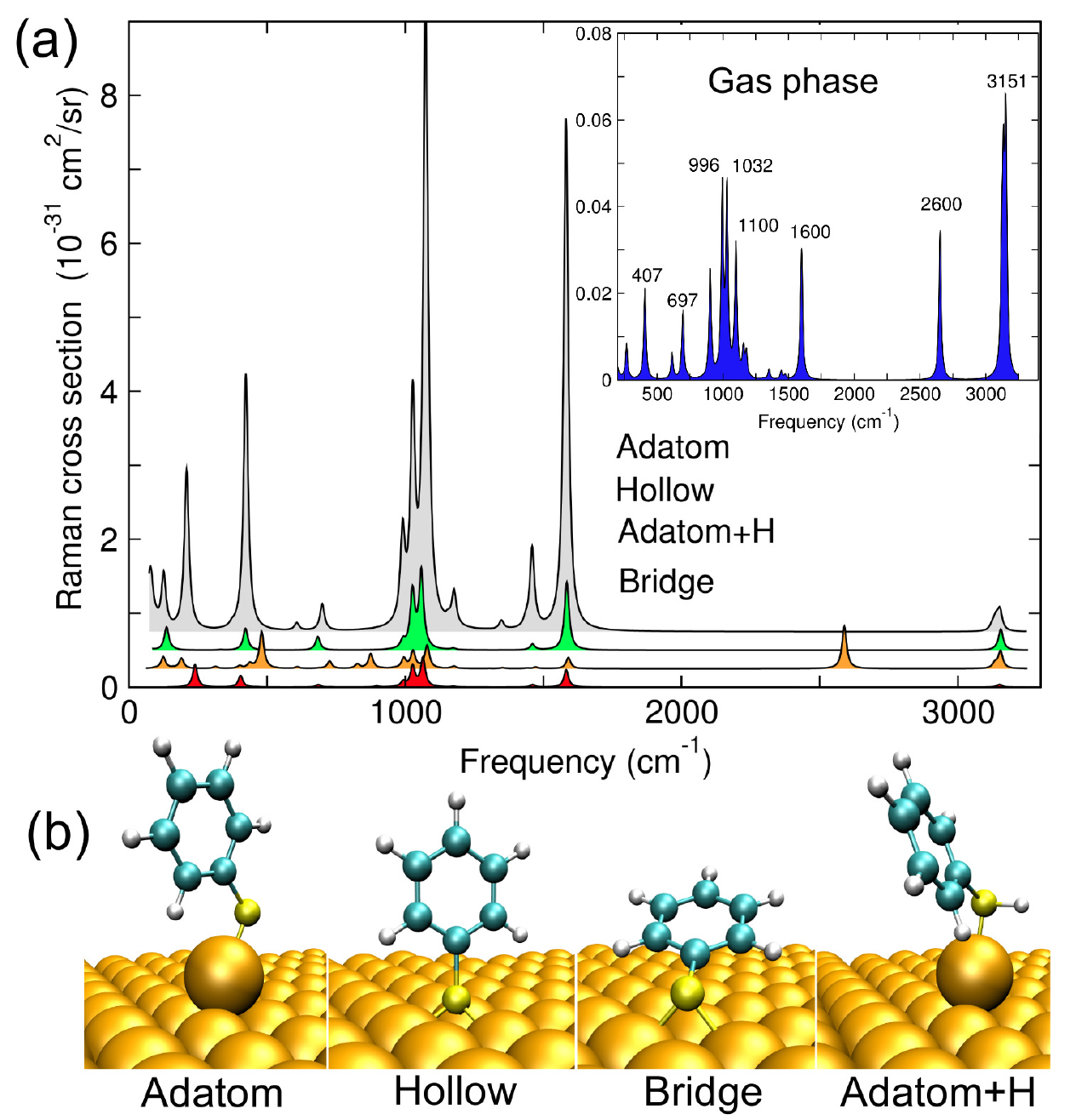}
 \caption{\label{f1}
(A) Calculated Raman spectra (from $R_{zz}$ components only) for four different
binding geometries: adatom, hollow, adatom+H, and bridge. Inset shows
our calculations of the orientationally-averaged gas-phase Raman
spectrum of benzene thiol. All spectra are broadened
with a Lorentzian of 10 \cm\ width. (B) Binding geometries. 
} 
\end{figure}
%

DFT calculations are performed using the Vienna Ab-initio Simulations
Package (VASP) and within a generalized gradient
approximation\cite{PBE,kresse:1995}. We model the BT adsorbate-Au substrate system with
an ordered monolayer of one BT molecule per nm$^2$ bonded to a flat
periodic Au(111) slab. Our supercell consists of 5 atomic layers of Au
stacked along [111] with 16 atoms per layer, with 30 \AA\ of vacuum. The
forces of the three upper layers and molecule are well converged to
less than 1 meV/\AA. The in-plane lattice parameters are kept fixed to
their computed Au \textit{fcc} bulk value of 4.17 \AA. A 400 eV plane-wave cutoff and 2x2x1
Monkhorst-Pack k-point mesh is used for calculations involving Au
slabs. 
Four binding geometries are considered, as shown in Fig. 1b: 
 \textit{fcc} hollow (E$_B$=0.219 eV), adatom (E$_B$=0.446 eV), hydrogenated
adatom (E$_B$=0.795 eV), and bridge (E$_B$=0.192 eV), where E$_B$ is the
calculated binding energy relative to a free Au surface and gas-phase
$C_6H_5SH$  in the dilute limit.
The gas-phase BT molecule is simulated in the same large supercell as the
slab geometry, but using the $\Gamma$ point only.

 Static Raman tensors are
constructed mode-by-mode using a finite-differences approach, in two
steps. First, the dynamical matrix of the system is generated by
displacing each atom along each Cartesian direction by 0.03 \AA. 
Vibrational frequencies and corresponding phonon eigenvectors are
obtained by diagonalization of a truncated dynamical matrix treating
only BT atoms, and Au atoms directly bonded with sulfur. (The
remainder of Au atoms is treated in an infinite mass approximation.)
Second, we compute the static polarizability within a second-order
finite-difference expression using a saw-tooth potential with a
gradient of 1 mV/\AA, and compute its derivative as a function of the
amplitude for each vibrational eigenmode. (See Supplementary
Information for additional details.) Throughout the paper, all modes
are labeled with gas-phase frequencies for simplicity. 
%
\begin{figure}
\includegraphics*[width=6.0cm]{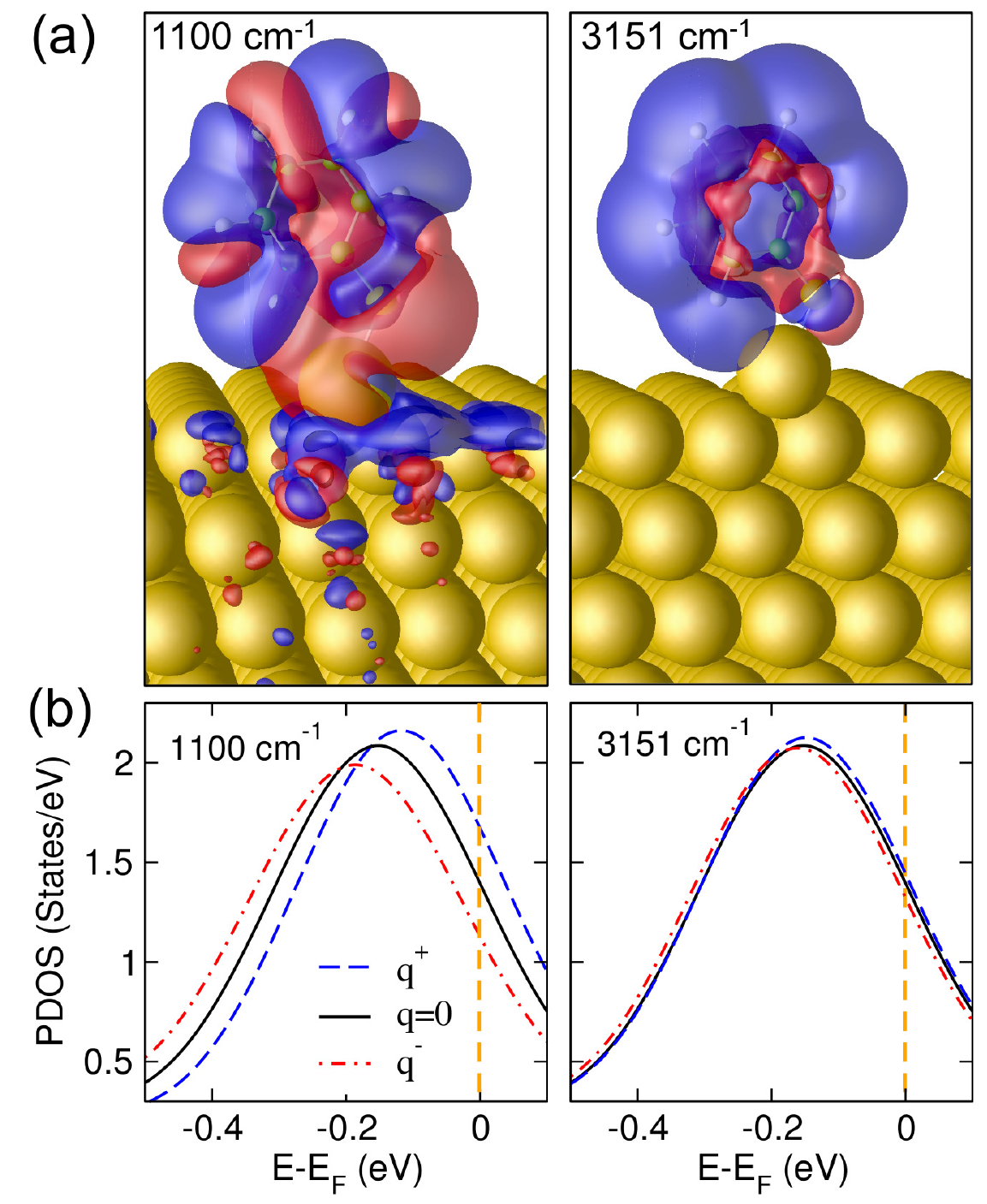}
 \caption{\label{f2}
(A) Computed isosurfaces (at 1.65 x 10$^{-4}$ e/\AA${^3}$) of charge density induced by freezing-in
small amplitudes (0.1 \AA) of the 1100 \cm\ and 3151 \cm\
modes. For both modes, gold atoms are stationary, and charge rearrangement in the
substrate is induced by the molecule. (B) Partial densities of states projected on the molecule
near the Au Fermi energy for three different amplitudes. By inspection of the wavefunctions,
the peak is the BT HOMO.} 
\end{figure}
%

In Fig. 1a, we report Raman cross sections
calculated from the dominant (non-evanescent) component of the Raman
tensor, $R_{zz}$, where $z$ is the normal to the surface. Prominent Raman peaks
in the computed spectra agree well with experiment\cite{Guieu}. Upon adsorption,
BT vibrational frequencies are altered on average by 10-20 cm$^{-1}$, in
agreement with experiments (Supplementary Information). From Fig. 1a,
the presence of the Au substrate enhances Raman cross sections for
some modes more than for others. The intensities of modes with larger
enhancements exhibit a stronger dependence on binding site. For
example, while the 1100 \cm\ mode varies by more than two orders of
magnitude across the different sites considered, the 3151 \cm\ mode
remains relatively unaltered and comparable to the gas-phase
intensity. 
%
\begin{figure}
\includegraphics*[width=8.0cm]{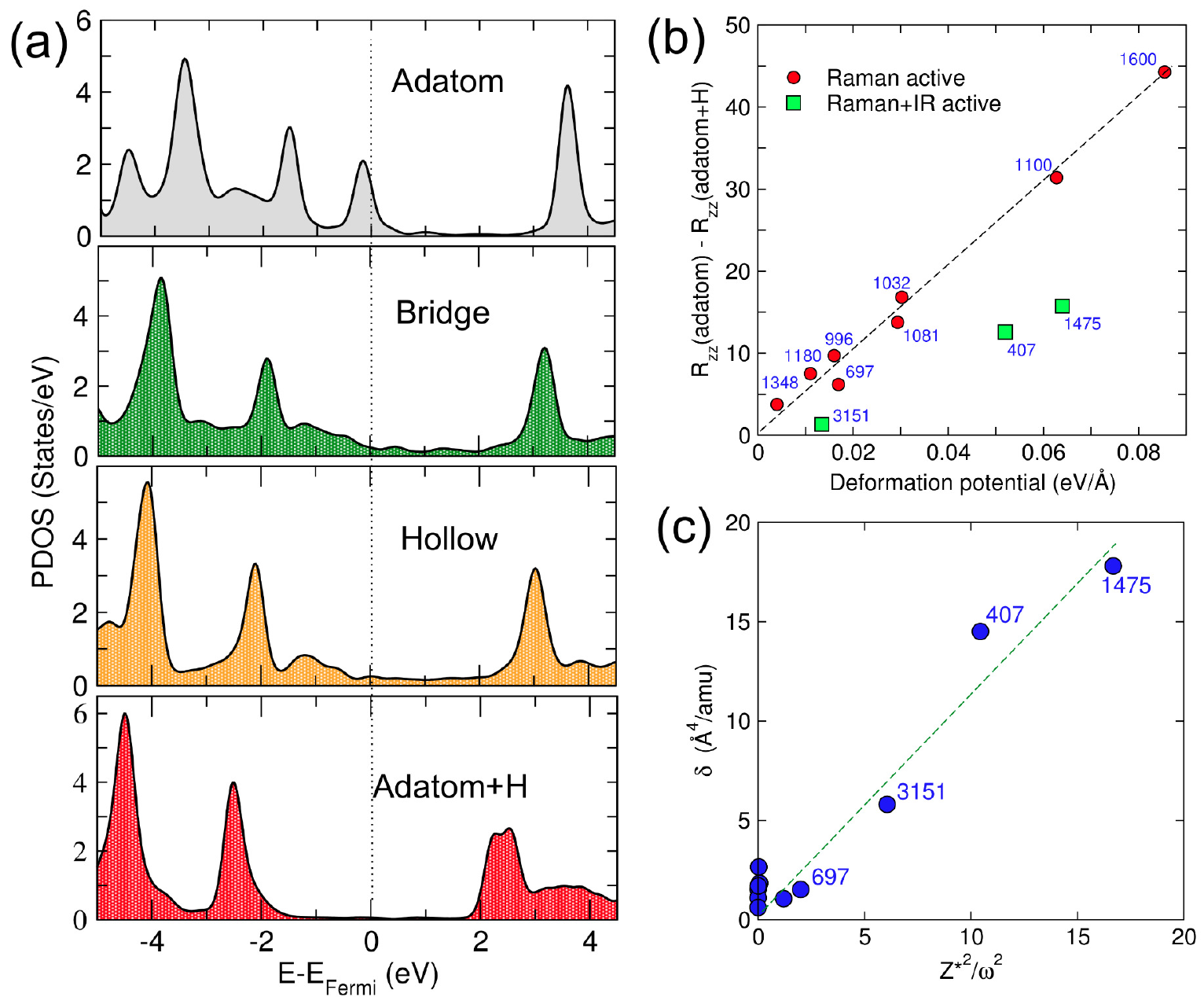}
 \caption{\label{f3}
(a) Partial densities of states projected on the molecule for
different binding sites shown about the Au
Fermi energy. (b) Difference $R_{zz}^{Adatom}-R_{zz}^{Adatom+H}$
plotted versus $\partial \omega_H / \partial Q_n$. (c)  The deviation $\delta$ of these modes from linear trend in (b) plotted versus their contribution to the lattice
susceptibility, which is proportional to the square of the ratio of the mode dynamical charge and its frequency.}
\end{figure}
%

In Fig. 2a, we compare the computed charge density change induced by
each mode,  $\delta \rho(Q_n) =
\rho(Q_n)-\rho(0)$, where $Q_n$ is the vibrational
eigenvector. We find that the 1100 \cm\ mode induces significant
charge redistribution within the Au surface, while the 3151 \cm\ mode
does not. This behavior is entirely consistent with mode-induced
changes observed in the DFT electronic structure (Fig.2b): The peak
Kohn-Sham HOMO energy E$_{HOMO}$ shifts noticeably relative to the Fermi
level E$_F$ with the 1100 \cm\ mode, whereas the 3151 \cm\ mode leaves
the HOMO peak unchanged. Examining all modes, we find that modes that
show larger enhancements induce a larger polarization response in the
substrate (and larger shift in E$_F$-E$_{HOMO}$). 

To rationalize our DFT results, we compare a model expression for the
static polarizability and Raman tensor with our more rigorous
first-principles calculations. We consider a single (dominant) term in
an approximate single-particle form for the ground state electronic
polarizability, one that includes just a lone virtual transition
between the HOMO, $\varphi_{HOMO}$ , and a metallic state at the Au
Fermi level, $\phi_{Au}$. Within this two-state approximation, an adsorbate-metal interfacial
contribution to the Raman tensor R$_{BT-Au}$ can be expressed as\cite{Jensen2008,Morton} 
\begin{equation}
 R_{BT-Au}=\frac{\partial \alpha_{BT-Au}}{\partial Q_n}\approx
 \frac{1}{\omega_H}
\frac{\partial M_i^* M_j}{\partial Q_n}-\frac{M_i^* M_j}{\omega_H^2}
\frac{\partial \omega_H}{\partial Q_n}
\end{equation}
where $M_i = \langle \varphi_{HOMO} \mid \mu_i \mid \phi_{Au} \rangle$
is a dipole matrix element in the Cartesian direction $i$, and $\omega
= E_{Fermi} - E_{HOMO}$
is the difference between their eigenvalues. For modes that are not
also IR-active, terms involving derivatives of induced dipole moments
can be neglected, and Eq. (1) reduces to a term proportional to
$\displaystyle -\frac{1}{\omega_H^2}\frac{\partial \omega_H}{\partial Q_n}$.  As
noted previously\cite{Morton}, the $\omega_H^2$  factor in the denominator is consistent with
a sensitivity of the overall spectral enhancement to the alignment of
the frontier molecular electronic orbital with the metal Fermi
energy. Indeed, this factor can be used to rationalize the binding
site dependence of enhancements shown in Fig. 1a, and corresponding
partial densities of states shown on Fig.3a. Calculated enhancements
on the adatom site, for which the HOMO is close to E$_F$, are
significantly stronger than on other sites, where this level is
broadened and further away from E$_F$. However, this $\omega_H^2$  factor is the same
for all modes, and cannot explain the mode-dependence of CE computed
in Fig. 1a. 

 The strong modification of the BT Raman spectra by the Au substrate
 can be explained via computation of a deformation potential,
$\partial \omega_H / \partial Q_n$, for
 each mode, i.e. the change in molecular electronic level alignment
 relative to the metallic Fermi level induced by a particular
 vibration mode. In Fig. 3b, we plot the difference $R_{zz}^{Adatom}-R_{zz}^{Adatom+H}$  against 
 $\partial \omega_H / \partial Q_n$. Because the Kohn-Sham HOMO level of the
 adatom+H geometry is much further from the Fermi energy than for the
 adatom binding site (see Fig. 3a), $\omega_H$ is large, and the adatom+H
 geometry has a negligible interfacial contribution. Thus, to an
 excellent approximation, taking the difference $R_{zz}^{Adatom}-R_{zz}^{Adatom+H}$  removes
 intramolecular contributions to the Raman tensor unrelated to
 Eq. (1). Indeed, in Fig. 3b, we find a remarkable correlation between
 the interface contribution to the deformation potential and
 enhancement for most of the vibrational modes. The modes that deviate
 from the linear trend at 407 \cm\ (Au-S stretch), 1475 \cm\ (phenyl
 ring stretch), and 3151 \cm\ (C-H stretch) have significant IR
 activity. In the static limit used here, polarization induced by
 these IR active modes screens the electric field experienced by the
 molecule, leading to a reduction in their Raman cross sections\cite{Cardona} by a
 factor proportional to $\left( Z^*_n\right)^2 / \omega_n^2$, with the
 mode dynamical charge given by $Z^*_n =\Delta \mu / \Delta Q_n$
 and where $\Delta \mu$  is the mode-induced interfacial dipole moment. In
 Fig. 3c, we show the deviation $\delta$ of the IR-active ``outlier'' modes
 from the linear trend observed for non-IR active modes versus $\left(
   Z^*_n\right)^2 / \omega_n^2$. The
 correlation between $\delta$ and  $\left( Z^*_n\right)^2 /
 \omega_n^2$  confirms that modes with larger
 contributions to the screening lead to greater deviations from the
 two-state model (Fig. 3b). We note however, that although important
 in our static calculations, this screening effect will have impact
 only below infrared frequencies, and thus will be inconsequential for
 typical probe frequencies, where the local fields will vary too
 rapidly for the IR-active vibrations to respond. 

With this information, we can now connect the strong modification of
Raman spectra by substrates to mode-specific changes of the electronic
structure of the metal-adsorbate interface: modes with the largest
interfacial contribution to the change in polarizability, as
quantified through a deformation potential, result in the most
substantial chemical enhancements. For BT, these modes are those that
break the conjugation of the HOMO (Supplementary
Information). Fig. 2a shows vividly how the 1100 \cm\ mode breaks the
resonant character of the carbon ring, while the 3151 \cm\ mode leaves
the $\pi$-symmetry of the electrons on the phenyl ring intact. This
suggests that for future adsorbates, the nature of modes with the
largest CE might be intuitively rationalized \textit{a priori}. 

To validate our theory of chemical enhancement, we compare with
experimental Raman and SERS measurements of BT on rough Au substrates
(Fig. 4a,b).  SERS substrates, consisting of roughened SiGe
surfaces (Fig.4a) coated with 30 nm of Au, were incubated in 3mM
BT solution (Sigma Aldrich W361607) in methanol overnight,
then gently rinsed with methanol and dried by nitrogen gas. Raman
(neat solution) and SERS spectra were collected at two wavelengths,
632.8 nm and 785 nm, using an inverted microscope set-up coupled to a
spectrometer (Acton SpectraPro 2300i) equipped with a
liquid-nitrogen-cooled charge-coupled device camera.
To eliminate uncertainties associated with
the number of molecular analytes in comparing Raman and SERS
intensities, we normalize ratios of Raman and SERS spectra peak
heights to the 996 \cm\ mode, which has only a modest enhancement from
our calculations. We note that absolute CEs can be obtained if we
normalize to a mode with zero deformation potential,  $\partial \omega_H / \partial Q_n$,  see mode 1348
\cm\ in Fig.3b, for example. As the 1348 \cm\ mode is not easily
observed experimentally, we use 996 \cm\ in this case. Assuming the
electromagnetic enhancement is the same for all modes, this relative
enhancement will reflect CE. This assumption is acceptable for modes
within a few hundred wavenumbers of the 996 \cm\ mode, based on the
width on relatively low Q-factor for localized plasmon resonance in Au
or Ag. 
%
\begin{figure}
\includegraphics*[width=8.0cm]{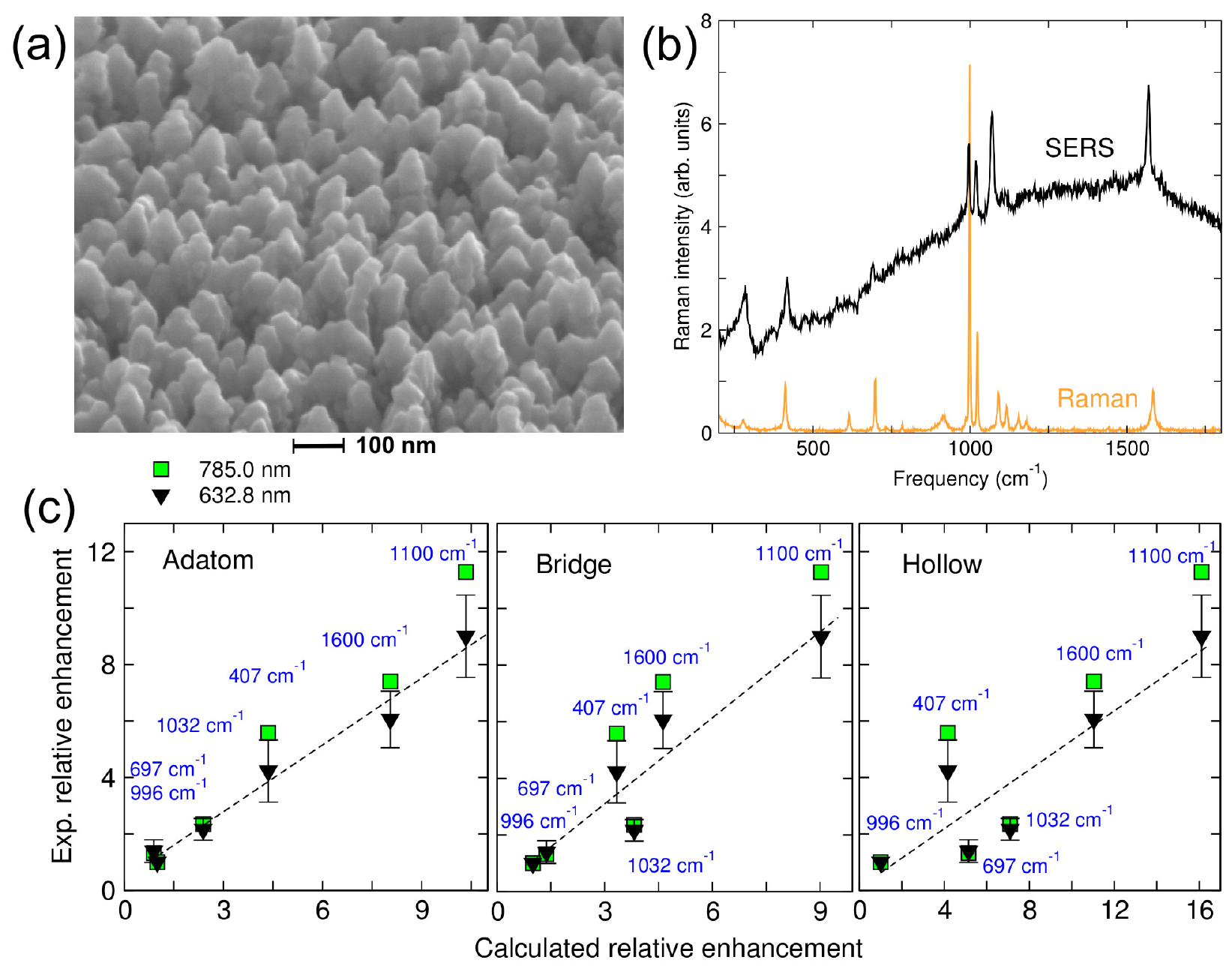}
 \caption{\label{f4}
(a) Au SERS substrate used in our measurements with BT. (b) Raman and
SERS BT spectra. (c) Calculated relative enhancements versus
experimental relative enhancements. Panels compare with calculations for
adatom, bridge and hollow sites.  
}
\end{figure}
%

In Fig.4c, we compare directly Raman cross sections  from Fig.1a
(computed from $R_{zz}$) to the averaged solution phase data from Fig.1a
(inset). Binding geometries for BT on flat Au surfaces are the subject
of debate in the literature; and for rough surfaces at room
temperature, a variety of adsorption sites will be available, and
binding geometries would be subject to thermal fluctuations\cite{Zenobi,Natelson}. By
comparing three very different, energy-minimized binding sites to the
experiment, we sample different possibilities for time-averaged
experimental binding morphologies. Surprisingly, all three geometries
show good correlation, with the adatom site showing perhaps the most
linear trend. (We note that SERS measurements do not observe the S-H
stretching mode at 2600 \cm\, indicating the loss of hydrogen at the
S-Au in the room-temperature experiments.) Importantly, the relative
theoretical enhancements, also normalized to the 996 \cm\ mode, are in
excellent quantitative agreement with the present experiments, as well
as others in literature taken from different substrates
\cite{Biggs,Aggarwal} (Supplementary Information).  


In summary, through our calculations and comparison with experiment,
we have demonstrated that the strong modification of Raman spectra by
the substrate is a chemical effect, largely independent of laser probe
frequency, and associated with the change in electronic structure of
the molecule by the metal substrate. The mode dependence of chemical
enhancement can be connected directly to interfacial contributions to
the deformation potential, a well-defined intrinsic property of each
vibrational mode and substrate. A new analysis of experimental SERS
data is introduced that allows for direct comparison of theoretical
calculations with experimental data. Comparing enhancements relative
to a particular mode, we find excellent agreement between theory and
experiment, indicating standard DFT approaches captures accurately and
quantitatively dominant contributions to CE, even in the static
limit. Additional support for our conclusions could be obtained by
inelastic electron tunneling measurements, which provide direct access
to interfacial contributions to the electron-vibron coupling. The
quantitative connection between the deformation potential and chemical
enhancement provides new opportunities for detection and control of
adsorbate-metal interactions through SERS. 

We thank L. Kronik, D. Prendergast, I. Tamblyn, and other colleagues
at Molecular Foundry and UC Berkeley for helpful discussions. This
work was supported by the AFOSR/DARPA Project BAA07-61 
``SERS S\&T Fundamentals'' under contract FA9550-08-1-0257, and the Molecular Foundry
through the Office of Science, Office of Basic Energy Sciences, of the U.S. Department of
Energy under Contract No. DE-AC02-05CH11231. Computational resources
were provided by DOE (LBNL Lawrencium, NERSC Franklin) and DOD (HPCMP
ARL MJM).

\bibliographystyle{apsrev4}
\bibliography{ref}

%

\end{document}